\documentclass[regular]{jfm}

\usepackage{graphicx}
\usepackage{newtxtext}
\usepackage{newtxmath}
\usepackage{natbib}
\usepackage{hyperref}
\usepackage{color}
\hypersetup{
    colorlinks = true,
    urlcolor   = blue,
    citecolor  = blue,
    linkcolor  = blue
}

\shorttitle{Settling Inertial Particles in Taylor-Couette Turbulence}
\shortauthor{H. Jiang, Z.-M. Lu, Yuan Ma and K.L. Chong}
\title{Role of settling inertial particles in modulating flow structures and drag in Taylor–Couette turbulence}

\author{
  Hao Jiang\aff{1},
  Zhi-Ming Lu\aff{1,2},
  Yuan Ma\aff{3},
	\and Kai Leong Chong \aff{1,2} \corresp{\email{klchong@shu.edu.cn}}
}

\affiliation{\aff{1} Shanghai Key Laboratory of Mechanics in Energy Engineering, Shanghai Institute of Applied Mathematics and Mechanics, School of Mechanics and Engineering Science, Shanghai University, Shanghai, 200072, China
\aff{2} Shanghai Institute of Aircraft Mechanics and Control, Zhangwu Road, Shanghai, 200092, China
\aff{3} Department of Civil and Environmental Engineering, The Hong Kong Polytechnic University, Hong Kong SAR, PR China
}

\begin{document}

\maketitle

\begin{abstract}
The modulation of drag through dispersed phases in wall turbulence has been a longstanding focus. This study examines the effects of particle Stokes number ($St$) and Froude number ($Fr$) on drag modulation in turbulent Taylor–Couette (TC) flow, using a two-way coupled Eulerian–Lagrangian approach with Reynolds number $Re_i = r_i \omega_i d/\nu$ fixed at 3500. Here, $St$ characterizes the particle’s inertia relative to the flow time scale, while $Fr$ describes the balance between gravitational settling and inertial forces in the flow. For light particles (small $St$), drag reduction is observed in the TC system, exhibiting a non-monotonic dependence on $Fr$. In specific, drag reduction initially increases and then decreases with stronger influence of gravitational settling (characterized by inverse of $Fr$), indicating the presence of an optimal $Fr$ for maximum drag reduction. For heavy particles, similar non-monotonic trend can also be observed, but significant drag enhancement is resulted at large $Fr^{-1}$. 

We further elucidate the role of settling particles in modulating the flow structure in TC by decomposing the advective flux into contributions from coherent Taylor vortices and background turbulent fluctuations. At moderate effects of particle inertia and gravitational settling, particles suppress the coherence of Taylor vortices which remarkably reduces angular velocity transport and thus leads to drag reduction. However, with increasing influence of particle inertia and gravitational settling, the flow undergoes abrupt change. Rapidly settling particles disrupt the Taylor vortices, shifting the bulk flow from a vortex-dominated regime to one characterized by particle-induced turbulence. With the dominance by particle-induced turbulence, velocity plumes---initially transported by small-scale G{\"{o}}rtler vortices near the cylinder wall and large-scale Taylor vortices in bulk region---are instead carried into the bulk by turbulent fluctuations driven by the settling particles. As a result, angular velocity transport is enhanced, leading to enhanced drag. These findings offer new insights for tailoring drag in industrial applications involving dispersed phases in wall-bounded turbulent flows.

\end{abstract}


\section{Introduction}\label{Introduction}

Multiphase flows are prevalent in both natural and engineering systems, where interactions between the dispersed and carrier phases give rise to a wide range of complex and nonlinear phenomena \citep{sippola2018experimental,carlotti2021experimental,brandt2022particle,zhang2023unveiling}. In particular, the presence of dispersed phases—such as solid particles, liquid droplets, or gas bubbles—can substantially alter the behavior of wall-bounded turbulence \citep{balachandar2010turbulent,russo2014water,mathai2020bubbly,bragg2021settling}. Depending on the physical properties and dynamic behavior of the dispersed phase, such interactions may lead to either drag reduction or enhancement, making this topic of both fundamental interest and practical relevance \citep{lu2005effect,zhang2020euler,kim2021direct,wang2022finite,ni2024deformation}. A deeper understanding of these mechanisms offers the potential to improve energy efficiency and flow control strategies in a variety of industrial applications.

Among canonical configurations for studying multiphase wall turbulence, Taylor-Couette (TC) flow—characterized by fluid motion between two concentric rotating cylinders—offers a well-controlled environment for investigating the underlying physics \citep{taylor1923viii,eckhardt2007torque,huisman2013logarithmic,grossmann2016high}. The interplay between centrifugal forces and viscosity in TC flow gives rise to rich dynamics, most notably the formation of Taylor vortices: large-scale, axisymmetric, counter-rotating structures that dominate angular momentum transport in the bulk flow \citep{ostilla2013optimal,ostilla2014boundary,grossmann2016high}. These coherent structures play a central role in determining the global flow state, and any modulation of their stability or strength can significantly alter flow features such as the velocity profile, turbulence intensity, and overall torque scaling.

Recent studies have shown that the introduction of a dispersed phase—such as particles or bubbles—into the TC system can modulate both the large-scale flow structure and the near-wall boundary layer, thereby affecting the angular momentum transport and flow regime transitions \citep{spandan2016drag,dash2020particle,baroudi2023taylor}. Among these, the drag modulation induced by buoyant bubbles has been particularly well studied. For instance, \citet{murai2005bubble} demonstrated experimentally that adding a small volume fraction (1\%) of rising bubbles to the TC flow could reduce drag by up to 36\%. Building on these observations, \citet{sugiyama2008microbubbly} conducted direct numerical simulations using an Eulerian–Lagrangian framework and confirmed that drag reduction arises from the ability of bubbles to generate axial forces that suppress Taylor vortices. Subsequent studies extended this framework to higher Reynolds numbers and found that the Froude number ($Fr$), which characterizes the relative strength of inertial to buoyant forces, plays a crucial role in controlling flow modulation \citep{spandan2016drag}. At low $Fr$, strong buoyancy causes bubbles to rise rapidly, disrupting the coherent Taylor vortices and reducing drag. At high $Fr$, however, the buoyancy effect weakens and bubbles become trapped within the vortices, diminishing their influence on the flow.

In rotating flows such as Taylor-Couette, the radial motion of particles is primarily governed by the density ratio between the particles and the fluid ($\rho^*$). Particles with a density ratio less than unity ($\rho^* < 1$) experience centripetal forces that drive them inward \citep{chouippe2014numerical,bakhuis2018finite}, whereas particles with a density ratio greater than unity ($\rho^* > 1$) are subjected to centrifugal forces that propel them outward \citep{wereley1999inertial,clarke2025movement}. This behavior contrasts with that of bubbles, which, due to their significantly lower density, predominantly migrate inward under centripetal effects. In our prior work, we elucidated the mechanism underlying the radial distribution of particles in the TC system under conditions characterized by high density ratios \citep{Jiang2025spatial}. For neutrally buoyant particles, where the density ratio $\rho^* \approx 1$, the dynamics differ significantly as buoyancy and centrifugal effects become negligible. In Taylor-Couette flow, neutrally buoyant particles exhibit distinct behaviors depending on the Reynolds number. At low $Re$, neutrally buoyant particles undergo inertial migration that influences flow transitions—destabilizing laminar states and stabilizing vortex flows \citep{majji2018inertial,baroudi2020effect}. At high $Re$, neutrally buoyant particles generally increase drag, with the extent of modification strongly affected by their shape and radial distribution, particularly near-wall clustering \citep{wang_yi_jiang_sun_2022,assen2022strong}.

While the effect of buoyant bubbles has been relatively well explored, the role of inertial particles—especially under the influence of gravitational settling—remains less well understood in the context of TC flow. Several studies have highlighted the importance of particle inertia in drag modulation in wall-bounded turbulence \citep{wang2021turbulence,costa2021near,gao2024drag}, demonstrating that particles can alter turbulence production, energy dissipation, and near-wall dynamics depending on their Stokes number. However, how the interplay between particle inertia and effective gravity (characterized by the Froude number) influences the global and local transport properties in TC systems is still an open question.

Addressing this knowledge gap is crucial for advancing our understanding of particle-laden wall turbulence and for designing efficient particle-based strategies for flow control. In this study, we investigate the combined effects of particle Stokes number ($St$) and Froude number ($Fr$) on drag modulation in turbulent TC flow using a two-way coupled Eulerian–Lagrangian approach. Our results reveal a non-trivial, non-monotonic relationship between drag and particle properties, and offer new insights into the mechanisms underlying structural transitions in particle-laden TC turbulence.

The remainder of the article is organized as follows: In \S \ref{sec:num}, we introduce the governing equations and numerical setup; In \S \ref{sec:flux}, we examine how the angular velocity flux and drag are modulated by settling particles, with a particular focus on their dependence on the particle Stokes number and Froude number; In \S \ref{sec:decomposition}, we decompose the flux contributions into advection, viscous diffusion, and particle-induced components to elucidate the mechanisms by which settling particles modulate drag in Taylor–Couette flow; In \S \ref{sec:flowstruc}, we investigate the modulation of flow structures by settling particles and demonstrate that particle addition suppresses the Taylor vortices. For heavy particles with sufficiently strong settling effects, we identify a transition to a regime of particle-induced turbulence; In \S \ref{sec:boundary}, we examine how the near-wall flow properties are influenced by the presence of particles. Finally, we provide conclusion and outlook in \S \ref{sec:conclusion}.

\section{Governing equations and numerical set-ups}\label{sec:num}

% \begin{figure*}
%   \centering
%   \includegraphics[width=0.4\linewidth]{sketch.pdf}
%   \caption{Sketch of the set-up of particle-laden Taylor-Couette flow.}
%   \label{fig:TC_SK}
% \end{figure*}

% The Taylor-Couette (TC) system consists of a viscous fluid confined in the gap between two rotating coaxially cylinders as shown in figure~\ref{fig:TC_SK}. The inner and outer cylinders rotate with angular velocities $\omega_i$ and $\omega_o$, respectively. $r_i$ and $r_o$ represent the radius of inner and outer cylinders, respectively. The gap width between the two cylinders is denoted as $d=r_o-r_i$. The geometric control parameters of TC system include aspect ratio $\Gamma = L/d$ and gap ratio $\eta = r_i / r_o$, where $L$ is the height of TC system. In this study, the outer cylinder is fixed, and the inner cylinder rotates with a constant angular velocity $\omega_i$. 

The Taylor-Couette (TC) system consists of a viscous fluid confined in the gap between two coaxially rotating cylinders. The inner and outer cylinders rotate with angular velocities $\omega_i$ and $\omega_o$, respectively. $r_i$ and $r_o$ represent the radius of inner and outer cylinders, respectively. The gap width between the two cylinders is denoted as $d=r_o-r_i$. The geometric control parameters of the TC system include aspect ratio $\Gamma = L/d$ and the gap ratio $\eta = r_i / r_o$, where $L$ is the height of TC system. In this study, the outer cylinder is fixed, and the inner cylinder rotates with a constant angular velocity $\omega_i$. 

The Euler-Lagrangian framework is used to carry out the inertial particles settling in turbulent TC flow. The fluid phase is described by the incompressible Navier-Stokes equations, while the particle phase is described by the dispersed point-particle. The coupling between the fluid and particle phases is achieved through the hydrodynamic force acting on the particles. The dimensionless governing equations for the fluid phase are the incompressible Navier-Stokes equations, given by
\begin{eqnarray}
    &\displaystyle{\nabla \cdot \pmb{u}=0,} \label{eq:Inc}
    \\
    &\displaystyle{\frac{\partial \pmb{u}}{\partial t} + (\pmb{u} \cdot \nabla) \pmb{u}=-\nabla p+\frac{1}{Re_i} \nabla^{2} \pmb{u} + \pmb{f}_p,} \label{eq:NS}
    \\
    &\displaystyle{\pmb{f}_p = -\frac{\rho_p}{\rho_f} V_p \sum_{i}^{N_p}(\frac{d\pmb{v}}{dt}-\frac{1}{Fr^2}\pmb{e}_z )\delta (\pmb{x}-\pmb{y}_i),} \label{eq:feedback}
\end{eqnarray}
where $Re_i = r_i \omega_i d/\nu$ is the Reynolds number based on the inner cylinder rotation, $\pmb{u}$ and $\pmb{v}$ represent the fluid and particle velocities, respectively. $\pmb{f}_p$ denotes the feedback force exerted by particles per unit mass of the fluid phase, $\rho_p$ and $\rho_f$ represent the particle and fluid densities, respectively. $V_p$ is the particle volume, $N_p$ is the number of particles, $Fr = \omega_i \sqrt{r_i/g}$ is the Froude number, $\pmb{e}_z$ is the unit vector in the axial direction, and $\pmb{y}_i$ denotes the instantaneous position of the $i$-th particle. The particle feedback force on the fluid is given by Eq. (\ref{eq:feedback}), where $\delta$ is the Dirac delta function, and $d\pmb{v}/dt$ is the acceleration of the particles. Equations (\ref{eq:Inc})-(\ref{eq:feedback}) are normalized by the inner cylinder rotation velocity $r_i \omega_i$ and gap width $d$. 

The particles dispersed in the fluid phase are tracked using a Lagrangian point-particle approach with the Stokes drag, added mass, Saffman-Mei lift, buoyancy and gravity force\citep{maxey1983equation,maxey1987gravitational,Gatignol_jmta_1983,Tsai_taml_2022}. The momentum equation for the particle phase is given by
\begin{align}
  \rho_p V_p \frac{d\pmb{v}}{dt} 
  &= \rho_p V_p \pmb{g}
  - C_D \frac{\pi d_p^2}{8} (\pmb{v} - \pmb{u}) |\pmb{v} - \pmb{u}|  + \rho_f V_p C_M \left( \frac{D\pmb{u}}{Dt} - \frac{D\pmb{v}}{Dt} \right)\notag \\
  &\quad
  + \rho_f V_p \left( \frac{D\pmb{u}}{Dt} - \pmb{g} \right)  - C_L \rho_f V_p (\pmb{v} - \pmb{u}) \times \pmb{\omega}, \label{eq:particle}
\end{align}
where the right-hand terms represent the gravitational force, Stokes drag force, added mass force, pressure gradient force and lift force, respectively. The particle density is denoted as $\rho_p$, $V_p = \pi d_p^3/6 $ is the particle volume, $d_p$ denotes the particle diameter. $\pmb{u}$ and $\pmb{\omega}$ are the velocity and vorticity of the fluid at the particle position, respectively. $\pmb{v}$ is the particle velocity, $\pmb{g}$ is the gravitational acceleration, while $C_D$, $C_M$ and $C_L$ represent the drag, added mass and lift coefficients, respectively. Using the characteristic velocity $r_i \omega_i$, characteristic length $r_i$ and fluid density $\rho_f$, the Eq.~(\ref{eq:particle}) can be normalized and simplified as
\begin{eqnarray}
  \frac{d\pmb{v}}{dt} = \frac{C_D}{St}(\pmb{u}-\pmb{v}) + \frac{D\pmb{u}}{Dt} + C_L (\pmb{u}-\pmb{v}) \times \pmb{\omega}+ \beta \frac{1}{Fr^2}\pmb{e}_z. \label{eq:normalized_particle}
\end{eqnarray}
Here, $St = {d_p^*}^2 Re_i / 12 (\beta+1)$ represents the particle Stokes number based on the inner cylinder rotation speed, $\beta = 2 (1-\rho^*)/(1+2\rho^*)$ is a parameter related to density ratio $\rho^*$. For the drag coefficient, the Shiller-Naumann relation $C_D = 1+0.15Re_p^{0.687}$ is used, which is valid for particle Reynolds number $Re_p = d_p|\pmb{v}-\pmb{u}|/\nu$ up to 1000 \citep{naumann1935drag}. The added mass coefficient is set to $C_M = 0.5$, and the lift coefficient is set to $C_L = 0.5$ \citep{spandan2016drag,sugiyama2008microbubbly}. 

Direct numerical simulations (DNS) of the carrier phase are performed using a second-order accurate finite-difference method in cylindrical coordinates $( \pmb{e}_r, \pmb{e}_{\theta}, \pmb{e}_z)$ \citep{verzicco1996finite,ostilla2013optimal}  with uniform grid spacing in the azimuthal and axial directions and a non-uniform grid spacing using a clipped Chebyshev-type clustering method in the radial direction. To reduce computational costs, we use a rotational symmetry $n_{sym}$ of 6 and aspect ratio $\Gamma = L/d = 4$ in the simulated TC system. These parameters have been validated for simulation accuracy of two-phase TC flow in \citet{spandan2016drag}. The no-slip boundary condition is applied at the inner and outer cylinder walls, and the periodic boundary condition is applied in the azimuthal and axial directions for the carrier phase and dispersed phase. The complete elastic collisions have been used to calculate the particle-wall collisions. The fluid information at the particle location and the feedback force~(\ref{eq:feedback}) extrapolated from the particle location to grid nodes are both calculated by a tri-linear scheme. In order to be consistent with \citet{sugiyama2008microbubbly} and \citet{spandan2016drag}, we also fix the mean axial flow flux to be zero at each $z$ position. In this study, the particle volume fraction is fixed at $6 \times 10^{-5}$, and the Reynolds number of the carrier phase is fixed at $Re_i = 3500$. The particle Stokes numbers $St$ range from 0.088 to 0.876, and the Froude number $Fr$ varies from 0.16 to 2.56. For comparison, the cases with $Fr^{-1} = 0$ are also considered, where the gravitational effects are neglected. For convenience, $Fr^{-1}$ will be used to represent the magnitude of the effective gravity in the following sections. The details of the particle parameters are summarized in table~\ref{tab:particle setup}.

\begin{table}
  \setlength{\tabcolsep}{0.3cm}
  \begin{center}
\def~{\hphantom{0}}
  \begin{tabular}{lccccccccc}
    
      $Fr^{-1}$     & $\rho^*$                 & $St$                                 & $\beta$                  \\[3pt]
      0             & 50, 100, 200, 500      & 0.088, 0.176, 0.375, 0.876  & -0.97, -0.985, -0.993, -0.997        \\
      0.39          & 50, 100, 200, 500      & 0.088, 0.176, 0.375, 0.876  & -0.97, -0.985, -0.993, -0.997        \\
      0.78          & 50, 100, 200, 500      & 0.088, 0.176, 0.375, 0.876  & -0.97, -0.985, -0.993, -0.997        \\
      1.56          & 50, 100, 200, 500      & 0.088, 0.176, 0.375, 0.876  & -0.97, -0.985, -0.993, -0.997        \\
      2.34          & 50, 100, 200, 500      & 0.088, 0.176, 0.375, 0.876  & -0.97, -0.985, -0.993, -0.997        \\
      3.13          & 50, 100, 200, 500      & 0.088, 0.176, 0.375, 0.876  & -0.97, -0.985, -0.993, -0.997        \\
      4.17          & 50, 100, 200, 500      & 0.088, 0.176, 0.375, 0.876  & -0.97, -0.985, -0.993, -0.997        \\
      5.26          & 50, 100, 200, 500      & 0.088, 0.176, 0.375, 0.876  & -0.97, -0.985, -0.993, -0.997        \\
      6.25          & 50, 100, 200, 500      & 0.088, 0.176, 0.375, 0.876  & -0.97, -0.985, -0.993, -0.997        \\
  \end{tabular}
  \caption{Details of the settling particle parameters in the numerical simulations. Here,  $\rho^* = \rho_p/\rho_f$ is the density ratio between the particle and fluid, $St = {d_p^*}^2 Re_i / 12 (\beta+1)$ is the particle bulk Stokes number. For all cases,  the dimensionless particle diameter  $d_p^* = d_p / d$ and volume fraction $\phi_v = N_p V_p/V_f $ are fixed at 0.003 and $6 \times 10^{-5}$, respectively. The particle mass loading ratio $\phi_m = N_pm_p/m_f$  ranges from $3.1\times10^{-3}$ to $3.1\times10^{-2}$. For $Fr^{-1} = 0$, it represents the case where the gravitational effects can be neglected.}
  \label{tab:particle setup}
  \end{center}
\end{table}

\section{Angular velocity flux and drag modulation} \label{sec:flux}

In TC flow, the torque $T$ and the angular velocity flux $J^{\omega}$ are related by $T = 2\pi L J^{\omega}$ \citep{eckhardt2007torque}, and $J^{\omega}$ is conserved in the radial direction. $Nu_{\omega}$ represents the dimensionless angular velocity flux, where it is defined by $J^{\omega}/J_{\textrm{lam}}$ with the normalization by the flux in laminar flow. The net percentage drag reduction can be defined as following \citep{spandan2016drag}
\begin{gather} 
  \mathrm{DR}=\frac{\left\langle Nu_\omega\right\rangle_s-\left\langle Nu_\omega\right\rangle_t}{\left\langle Nu_\omega\right\rangle_s} \times 100\%=\frac{\left\langle C_f\right\rangle_s-\left\langle C_f\right\rangle_t}{\left\langle C_f\right\rangle_s} \times 100\%, \label{eq:DR}
\end{gather}
where $\left\langle \cdots \right\rangle_s$ and $\left\langle \cdots \right\rangle_t$ correspond to averaging for single-phase and two-phase systems, respectively. 

\begin{figure*}
  \centering
  \includegraphics[width=1\linewidth]{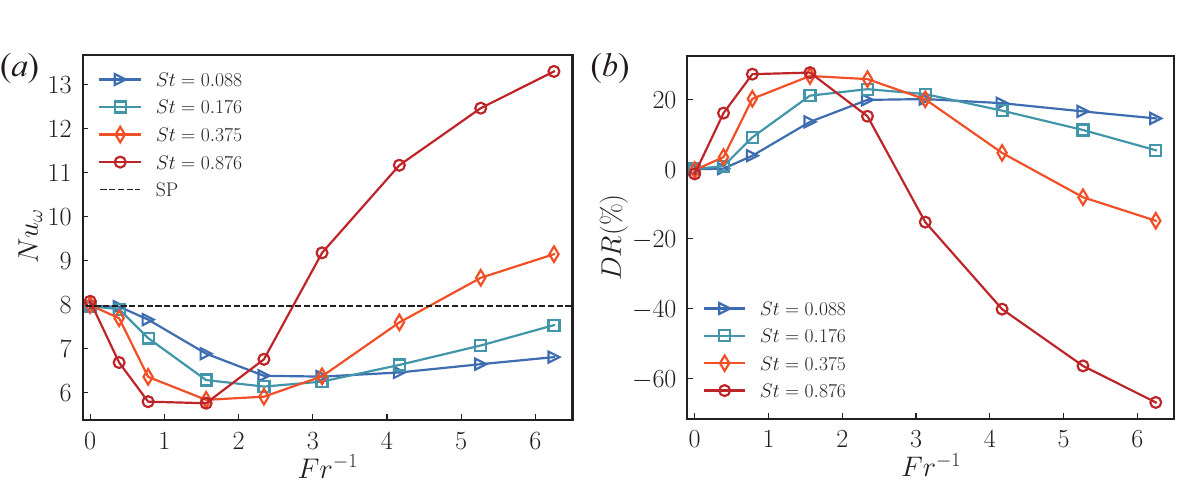}
  \caption{($a$) Nusselt number $Nu_{\omega}$ for two-phase Taylor-Couette (TC) flows under varying particle Stokes numbers ($St $) and inverse Froude numbers ($Fr^{-1}$), compared to the single-phase (SP) flow (horizontal dashed line) at $Re_i = 3500$. ($b$) The net percentage drag reduction for the two-phase TC flows, illustrating the influence of particle $St$ and $Fr^{-1}$.}
  \label{fig:Nu_DR}
\end{figure*}

We first present results on the influence of the settling particles on the drag reduction (DR) in TC flow. $Nu_\omega$ for two-phase TC flows is presented in figure~\ref{fig:Nu_DR}($a$), illustrating the influence of particle Stokes numbers ($St$) and inverse Froude numbers ($Fr^{-1}$) on the flow angular velocity transport and the corresponding drag reduction in figure~\ref{fig:Nu_DR}($b$). With varying strength of particle gravitational settling, $Nu_\omega$ exhibits non-monotonic variation. As $Fr^{-1}$ increases, there is more than 20\% drag reduction, where the amount of drag reduction reaches maximum value at certain optimal $Fr^{-1}$. For heavy particles ($St = 0.375, 0.876$), $Nu_{\omega}$ increases dramatically after the optimal point leading to drag enhancement in the TC flow for large $Fr^{-1}$ limit. The observed behaviour leads to the question on what leads to the drag reduction and the subsequent enhancement in drag.

\begin{figure*}
  \centering
  \includegraphics[width=1\linewidth]{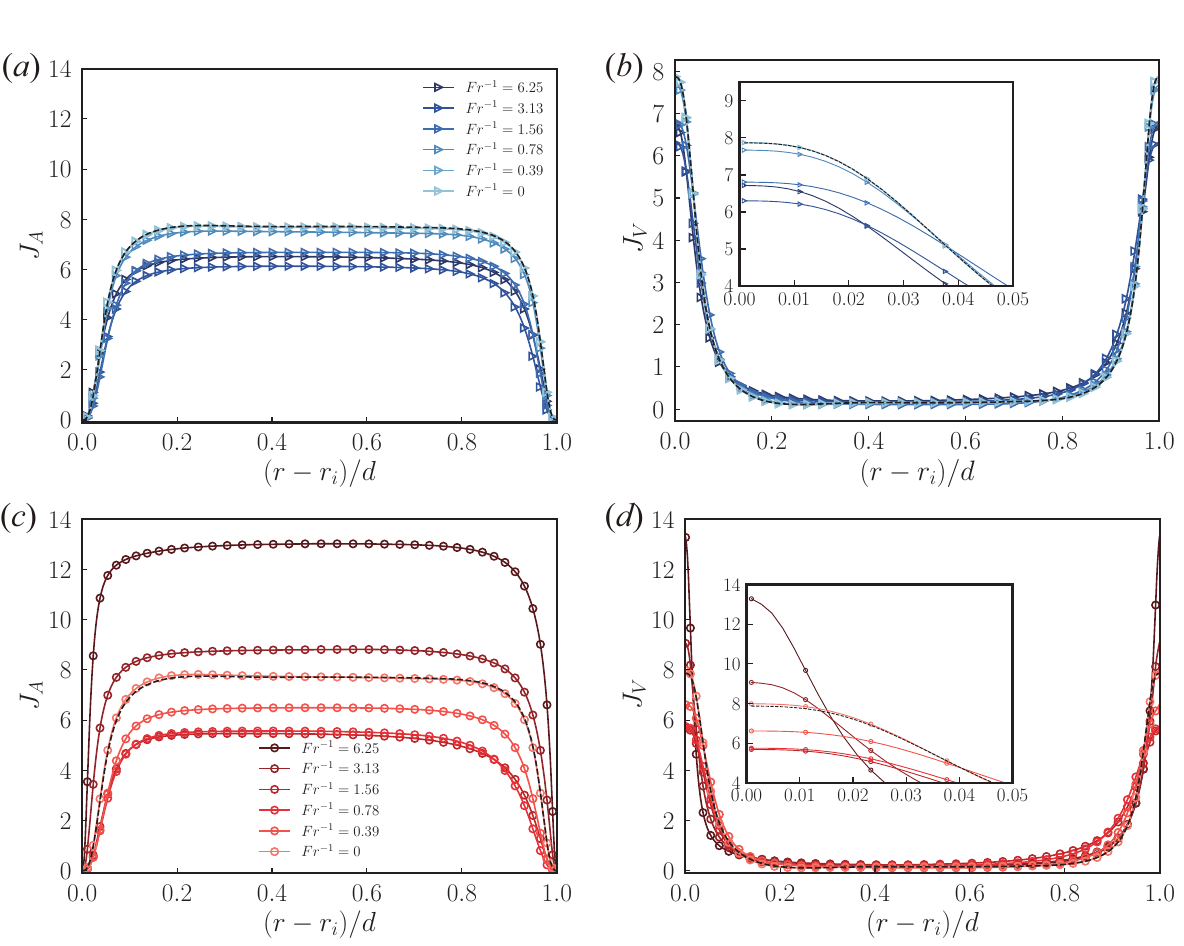}
  \caption{Dimensionless contributions of angular velocity flux for varying particle $St$ and $Fr^{-1}$, compared to the single-phase flow (dashed line). ($a$-$b$) The advection terms and viscous diffusion terms for $St = 0.088$; ($c$-$d$) the advection terms and viscous diffusion terms for $St = 0.876$.}
  \label{fig:All_J}
\end{figure*}

\section{Decomposition of flux contribution} \label{sec:decomposition}

To investigate the mechanism of drag modulation, we decompose the angular velocity flux into various parts. For the two-phase TC system, the angular velocity flux $J^{\omega}$ must also account for the feedback force from the particles. Following the approach proposed in \citet{su2024numerical}, the angular velocity flux $J^{\omega}$ for two-phase TC flow can be decomposed into three parts
\begin{gather}
  J^{\omega} = J_A(r)+J_V(r)+J_P(r) = const., \label{eq:DecJ}
\end{gather}
where $J_A = r^3\left\langle u_r \omega\right\rangle_{\theta,z,t}$, $J_V = -r^3\nu\left\langle \partial_r \omega\right\rangle_{\theta,z,t}$ and $J_P = -\int_{r_i}^{r}\left\langle r^2f_{\theta}\right\rangle_{\theta,z,t}$ represent the contributions from the advection, viscous diffusion and particle feedback to the angular velocity flux, respectively. Here, $\left\langle \cdots\right\rangle_{\theta,z,t}$  denotes averaging over time and in the $\theta,z$ direction. 

The contributions of $J_A$, $J_V$ normalized by the corresponding values can be obtained from the laminar flow $J_{\textrm{lam}}$ in figure~\ref{fig:All_J}. Despite the addition of particles, the contribution of particle feedback term $J_P$ to the angular velocity flux remains below 1\% across all cases. Therefore, this term has been neglected in the subsequent analysis. In contrast, the advection term predominantly governs the bulk region, while viscous diffusion dominates within the boundary layers of the Taylor–Couette system, consistent with the findings of \citet{brauckmann2013direct}.

When particle settling is neglected (i.e., $Fr^{-1}=0$), the contributions to angular velocity flux exhibit minimal variation, indicating that in the absence of gravitational settling, particles have a negligible influence on the Taylor-Couette flow. For light particles ($St = 0.088$), the introduction of effective gravity leads to a reduction in both advective and viscous diffusion contributions within their respective dominant regions, consistent with the observed drag reduction. In contrast, for heavy particles ($St = 0.876$), these contributions initially decrease but subsequently increase with increasing effective gravity, eventually surpassing the single-phase values when $Fr^{-1}>3.13$.

\begin{figure*}
  \centering
  \includegraphics[width=1\linewidth]{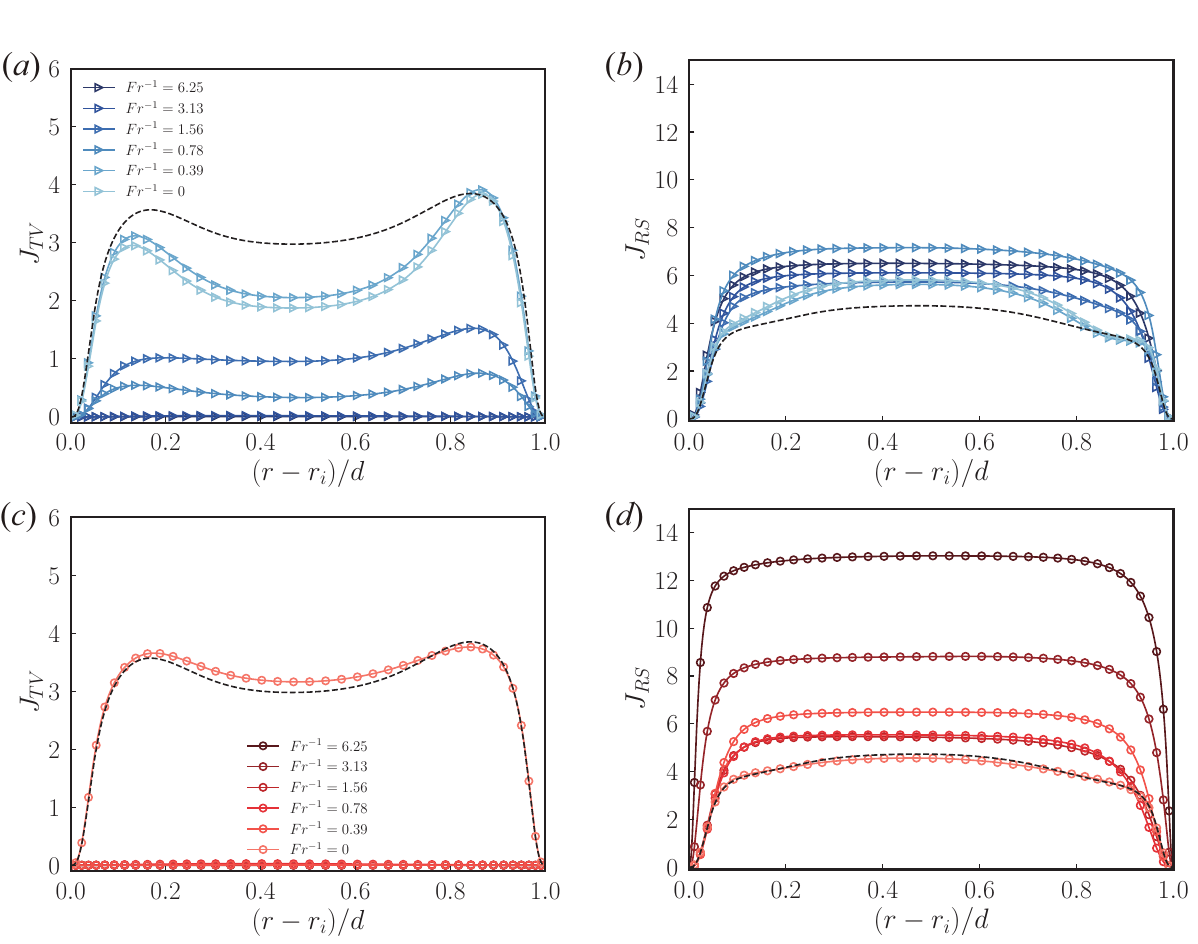}
  \caption{Dimensionless contributions of Taylor vortex and Reynolds stress as a function of radial position. ($a$) Taylor vortex contribution, $St=0.088$, ($b$) Reynolds stress contribution, $St=0.088$, ($c$) Taylor vortex contribution, $St=0.876$, ($d$) Reynolds stress contribution, $St=0.876$.} 
  \label{fig:T_R}
\end{figure*}

\section{Modulation of flow structure by settling particles} \label{sec:flowstruc}

Note that TC turbulence consists of both turbulent Taylor vortices and background fluctuations, it is meaningful to decompose the advection contribution of flux $J_A = r^3\left\langle u_r \omega\right\rangle_{\theta,z,t}$ into two components \citep{eckhardt2007torque,zhang2025global}
\begin{gather}
  r^3\left\langle u_r \omega\right\rangle_{\theta,z,t} = r^3\left\langle \left\langle u_r\right\rangle_{\theta,t} \left\langle \omega\right\rangle_{\theta,t} \right\rangle_{z}+ r^3\left\langle u_r^{\prime} \omega^{\prime} \right\rangle_{\theta,z,t}, \label{eq:JTV}
\end{gather}
where $\left\langle \left\langle u_r\right\rangle_{\theta,t} \left\langle \omega\right\rangle_{\theta,t} \right\rangle_{z}$ and $\left\langle u_r^{\prime} \omega^{\prime} \right\rangle_{\theta,z,t}$ represent the contribution of the Taylor vortex and Reynolds stress to the angular velocity flux, respectively. The dimensionless form of Taylor vortex contribution $J_{TV} = r^3\left\langle \left\langle u_r\right\rangle_{\theta,t} \left\langle\omega\right\rangle_{\theta,t} \right\rangle_z /J_\textrm{lam}$ and Reynolds stress contribution $J_{RS} = r^3\left\langle u_r^{\prime}\omega^{\prime} \right\rangle_{\theta,z,t}/J_\textrm{lam}$ are shown in figure~\ref{fig:T_R}. For the single-phase case, the magnitudes of $J_{TV}$ and $J_{RS}$ are approximately equal, indicating that Taylor vortices contribute about half of the angular velocity transport, which is consistent with the results of \citet{brauckmann2013direct}. 

%In the absence of gravitational effects ($Fr^{-1}=0$), the presence of light particles has a negligible net impact on the angular velocity flux, owing to a compensatory mechanism: the suppression of the Taylor vortex-induced flux $J_{TV}$ is offset by an enhancement in the contribution from background turbulent fluctuations in the bulk region. This insensitivity contrasts with the behavior observed for heavy particles, which exert only a minor influence on both $J_{TV}$  and the fluctuation-driven flux $J_{RS}$ under the same conditions. The underlying difference arises from the distinct spatial distributions and transport characteristics of the particles: 

%In the absence of gravitational effects, the light particles achieve a balance in $Nu_{\omega}$ by suppressing $J_{TV}$, whereas heavy particles have a slight influence on both $J_{TV}$ and $J_{RS}$. This is because light particles tend to be more uniformly distributed in the bulk region and are passively transported by the flow, with angular velocity being transferred from the continuous phase to the dispersed phase, thereby suppressing the mean Taylor vortices and enhancing the fluctuations in the bulk region. In contrast, due to centrifugal effects, heavy particles accumulate near the outer cylinder wall, leading to little influence on advection in the bulk region. As the gravitational effects increase, the Taylor vortex effect quickly diminishes, and the Reynolds stress increases significantly; this indicates that the transport of angular velocity is predominantly carried out through the turbulent fluctuations.

\begin{figure*}
  \centering
  \includegraphics[width=1\linewidth]{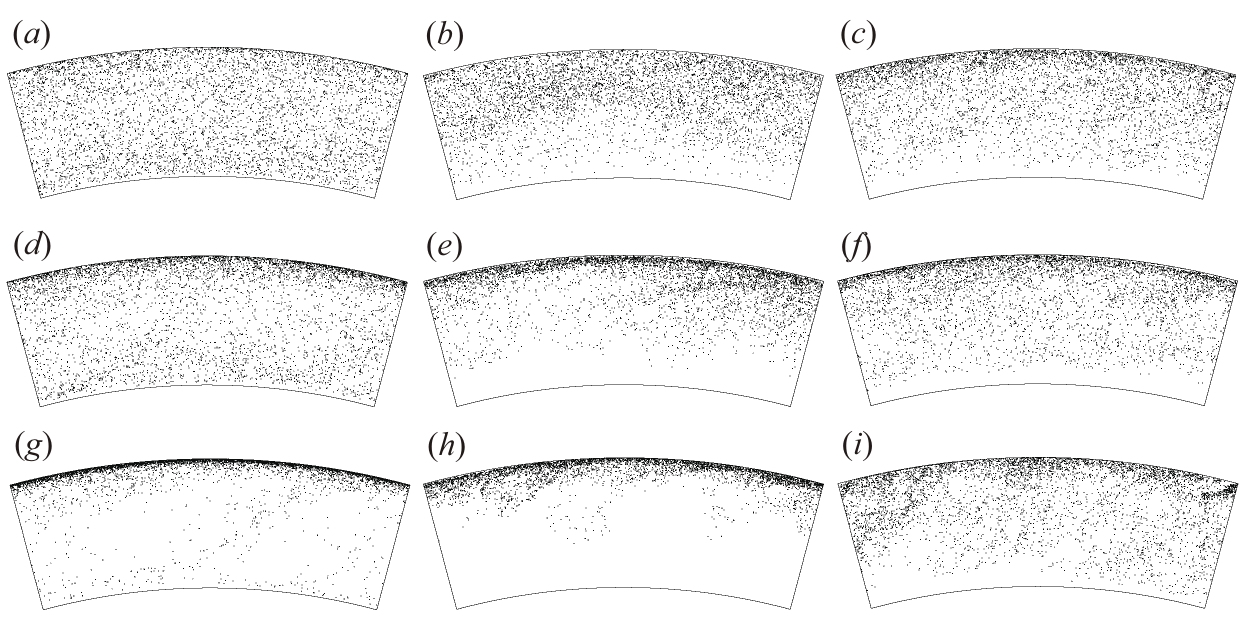}
  \caption{Top-view instantaneous snapshots of particle distribution for varying Stokes numbers, $St = 0.088$ ($a$–$c$), $St = 0.176$ ($d$–$f$), and $St = 0.876$ ($g$–$i$), and varying inverse Froude numbers, $Fr^{-1} = 0$ ($a$, $d$, $g$), $Fr^{-1} = 2.34$ ($b$, $e$, $h$), and $Fr^{-1} = 5.26$ ($c$, $f$, $i$). Only the angular range $\theta \in (0,\pi/6)$ is shown.}
  \label{fig:ppart}
\end{figure*}

At $Fr^{-1}=0$, the addition of light particles exerts minimal influence on the overall angular velocity flux, due to a balance between two competing effects: the suppression of the Taylor vortex contribution $J_{TV}$ and the simultaneous enhancement of flux from turbulent fluctuations $J_{RS}$ in the bulk region. This insensitivity differs notably from the case of heavy particles, which show only a weak influence on both $J_{TV}$ and $J_{RS}$ under the same conditions. The distinction arises from the differing spatial distributions and transport dynamics of the particles: light particles are more uniformly dispersed throughout the bulk and are passively carried by the flow, facilitating the transfer of angular momentum from the fluid to the particles (see figure \ref{fig:ppart}). This interaction suppresses the coherence of mean Taylor vortices while amplifying the intensity of background fluctuations. In contrast, heavy particles experience strong centrifugal forces, causing them to accumulate near the outer cylinder wall and interact less with the bulk flow, thereby exerting limited influence on advective transport. 

As the effect of gravitational settling increases ($Fr^{-1}$ rises), the contribution from coherent Taylor vortices progressively diminishes, becoming minimal in the case of heavy particles. In contrast, the contribution from background turbulent fluctuations increases significantly with rising $Fr^{-1}$, indicating a transition toward a flow regime dominated by particle-induced turbulence. %\textcolor{red}{This behavior contrasts with the findings of \citet{zhang2025global}, where polymer-induced drag reduction does not significantly alter the structure of Taylor vortices. In their case, drag reduction occurs primarily through viscoelastic damping of small-scale turbulence, while the large-scale vortical structures remain largely intact. In our study, however, gravitational settling fundamentally disrupts the formation and sustenance of Taylor vortices, demonstrating a distinct mechanism in which buoyancy-driven effects directly reshape the large-scale flow dynamics.}

\begin{figure*}
  \centering
  \includegraphics[width=1\linewidth]{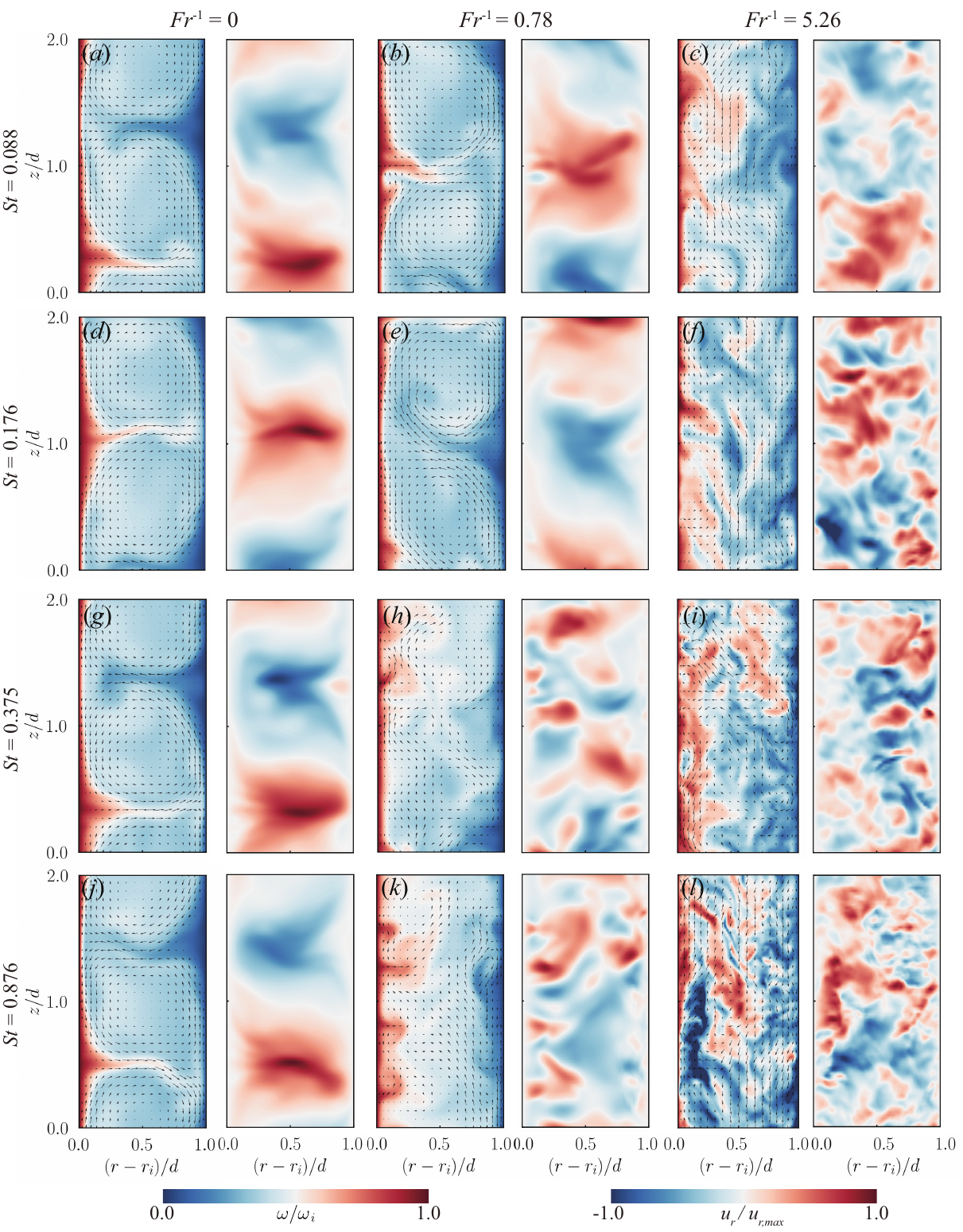}
  \caption{The ($r$-$z$) plane angular velocity contours ( left ) and radial velocity contours ( right ) for varying $St$, which are $(a-c)$ $St=0.088$,  $(d-f)$ $St=0.176$, $(g-i)$ $St=0.375$ and $(j-l)$ $St=0.876$, and $Fr^{-1}$, which are $(a,d,g,j)$ $Fr^{-1}=0$, $(b,e,h,k)$ $Fr^{-1}=0.78$ and $(c,f,i,l)$ $Fr^{-1}=5.26$. Only the axial domain $z \in (0,2)$ is shown.}
  \label{fig:thcut}
\end{figure*}

The vanishing of the Taylor vortex contribution ($J_{TV}$) does not imply the complete absence of Taylor vortices in the TC system. When the vortices exhibit axial migration or strong temporal fluctuations, the averaging operation can cause $J_{TV}$ to vanish despite the vortices still being present. To clarify the underlying flow states in the two-phase TC system, we present instantaneous snapshots of angular velocity and radial velocity in the radial–axial plane, as shown in figure~\ref{fig:thcut}. In the absence of settling effect, Taylor vortices are clearly visible in the bulk region, where they dominate the flow. At small values of $Fr^{-1}$, increasing particle inertia causes the edges of the Taylor vortices to become increasingly blurred, accompanied by axial oscillations. In this case, velocity plumes near the wall are primarily governed by small-scale G{\"{o}}rtler vortices \citep{DONG2007Direct,zhang2025global}. At $St = 0.176$ and $Fr^{-1} = 0.78$, Taylor rolls are still discernible in the instantaneous snapshots, although the contribution from $J_{TV}$ has already diminished.

For heavy particles ($St = 0.375, 0.876$), even a small effective gravity disrupts the Taylor vortices, as shown in figure~\ref{fig:thcut}($h,k$), leading to rapid suppression of the mean Taylor vortex structures and a concurrent increase in background turbulent fluctuations. However, this enhancement of turbulence is weaker than the suppression of the vortex contribution, resulting in significant drag reduction in the system. To further analyze this, we compute $\left\langle u_r\right\rangle_{\theta,t}$ in accordance with Eq. (\ref{eq:JTV}), and found that when the mean Taylor vortex effect is present, the averaged $\left\langle u_r\right\rangle_{\theta,t}$ field clearly exhibits the characteristic structure of Taylor vortices. In contrast, when the mean Taylor vortex contribution is negligible, the averaged $\left\langle u_r\right\rangle_{\theta,t}$ field is nearly zero, indicating that the flow in the bulk region is predominantly governed by turbulent fluctuations, as shown in figure~\ref{fig:thcut} for the large $Fr^{-1}$.

\begin{figure*}
  \centering
  \includegraphics[width=1\linewidth]{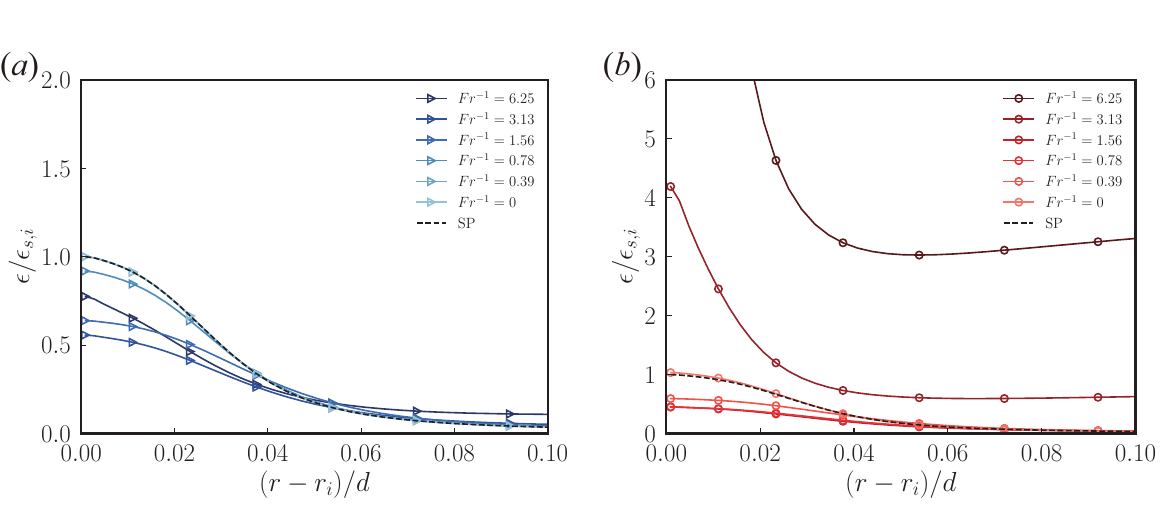}
  \caption{Viscous dissipation rate near the inner cylinder wall, normalized by the single-phase case $\epsilon/\epsilon_{s,i}$. ($a$) $St = 0.088$, ($b$) $St = 0.876$.}
  \label{fig:diss}
\end{figure*}

\begin{figure*}
  \centering
  \includegraphics[width=1\linewidth]{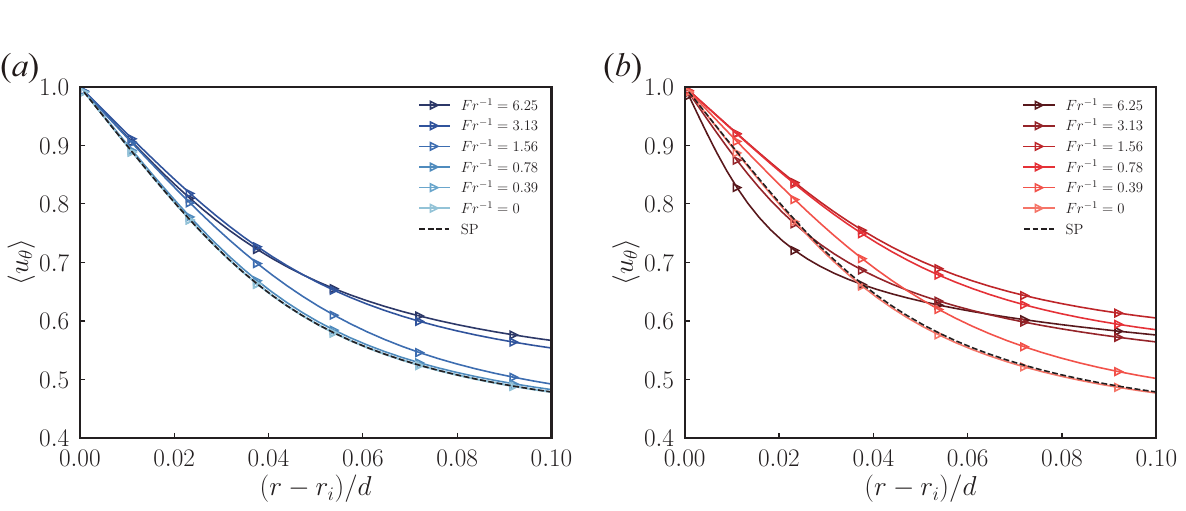}
  \caption{Azimuthal velocity profile near the inner cylinder wall. ($a$) $St = 0.088$, ($b$) $St = 0.876$.}
  \label{fig:uth}
\end{figure*}

\section{Near wall properties modulated by particles} \label{sec:boundary}
Next, we examine how changes in the flow structure lead to distinct near-wall properties, such as viscous dissipation rates. This analysis is essential, as drag reduction is typically associated with a decrease in viscous dissipation \citep{eckhardt2007torque,sugiyama2008microbubbly,spandan2016deformation}.  To understand how settling particles influence drag in the near-wall region, figure~\ref{fig:diss} presents the dissipation rate normalized by the single-phase dissipation at the inner cylinder wall, denoted as $\epsilon_{s,i}$. We compare the two-phase dissipation profiles for light particles ($St = 0.088$) and heavy particles ($St = 0.876$).

For light particles, the near-wall viscous dissipation is lower than in the single-phase case, consistent with the observed drag reduction. In contrast, for heavy particles under the drag enhanced cases, the near-wall dissipation is significantly elevated, with peak values reaching up to 30 times those of the single-phase system. To further assess how particles modulate the near-wall velocity field, we plot the azimuthal velocity profiles near the inner cylinder in figure~\ref{fig:uth}. Light particles are found to enhance the velocity close to the wall. In contrast, under strong effective gravity, heavy particles cause a sharp drop in the near-wall azimuthal velocity. This sharp change in velocity implies that the near-wall velocity gradient can be significantly altered by the presence of settling particles, suggesting that particles can effectively modify the boundary layer thickness.

\begin{figure*}
  \centering
  \includegraphics[width=0.6\linewidth]{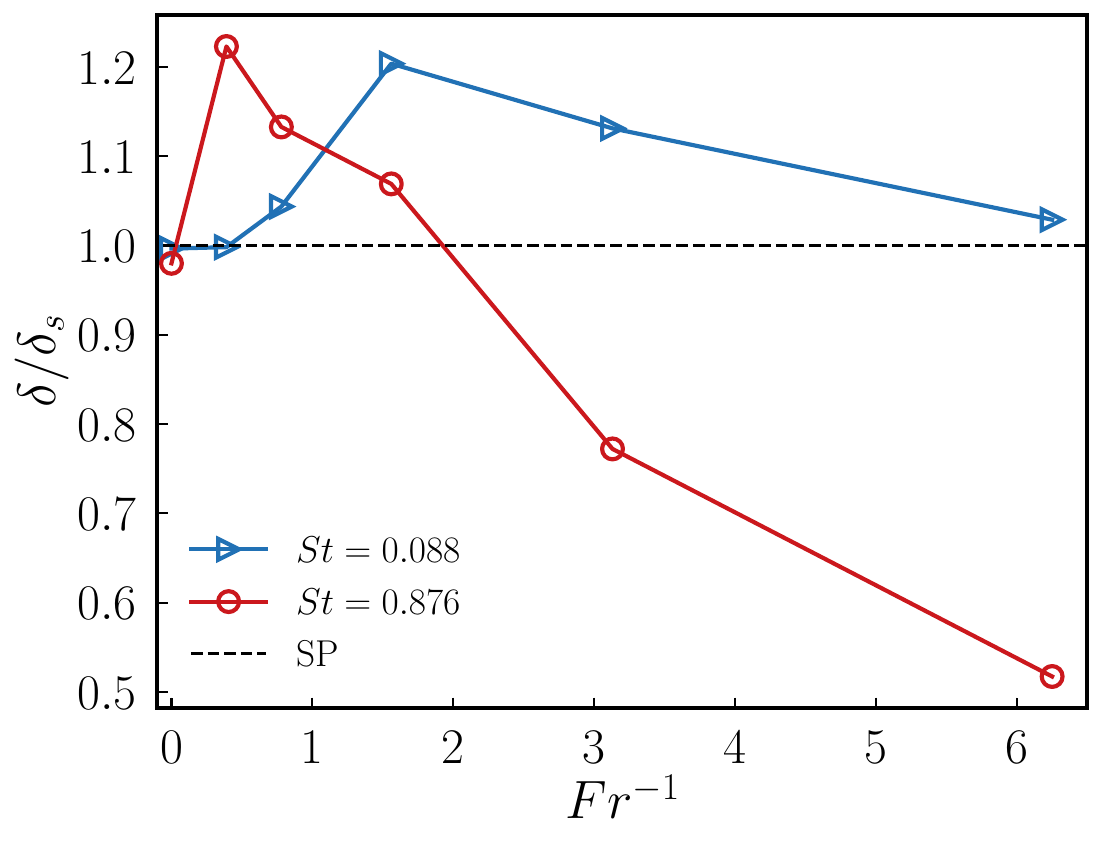}
  \caption{Inner boundary layer thickness normalized by the single-phase case.}
  \label{fig:BTK}
\end{figure*}

Finally, we estimated the inner velocity boundary layer thickness from the azimuthal velocity ($u_{\theta}$) profile, as shown in figure~\ref{fig:BTK}, following the method of \citet{ostilla2013optimal}. Our results reveal that the variation in boundary layer thickness closely correlates with the observed drag reduction: an increase in boundary layer thickness corresponds to reduced drag, while a thinner boundary layer is associated with increased drag. This trend suggests that particle-induced modifications to the near-wall shear and momentum transport play a key role in altering the boundary layer dynamics. Specifically, a thinner boundary layer intensifies velocity gradients, which can increase turbulent momentum transport and lead to drag enhancement. Conversely, a thicker boundary layer may reflect a damping of near-wall turbulence and a weakening of Taylor vortices, thereby reducing the exchange of angular momentum and contributing to drag reduction.

These results indicate that particle settling modulates drag not only directly by altering the large-scale coherent structures in the bulk region, but also indirectly by influencing energy dissipation and the boundary layer thickness near the inner cylinder wall—even in regions with relatively low particle number density.

\section{Conclusions and outlook}\label{sec:conclusion}
In summary, we have systematically studied how particle inertia and gravitational settling modulate drag in turbulent Taylor-Couette flow using a two-way coupled Eulerian-Lagrangian framework. By varying the Stokes number and Froude number, we revealed a non-monotonic behavior in drag modulation. Light particles (small $St$) suppress coherent Taylor vortices and reduce drag, while heavy, fast-settling particles enhance turbulent transport and ultimately increase drag.

A key finding is that drag modulation arises not directly from the instantaneous momentum exchange between particles and fluid, but rather from particle-induced reorganization of the flow topology. This reorganization manifests as a competition between coherent Taylor vortices and turbulence driven by particle dynamics. Through decomposition of the angular velocity flux, we quantified the contributions from advection, viscous diffusion, and Reynolds stress, thereby clarifying the mechanisms underlying the transition between drag-reducing and drag-enhancing regimes.

Further insight was gained by decomposing the advective flux into contributions from Taylor vortices and background turbulent fluctuations. At moderate particle inertia and gravitational settling (characterized by $Fr^{-1}$), particles weaken the coherence of Taylor vortices, thereby reducing angular velocity flux transport and leading to drag reduction. As gravitational settling becomes stronger, the flow undergoes a structural transition: rapidly settling particles disrupt the Taylor vortices and promote a shift from a vortex-dominated regime to one governed by particle-induced turbulence. This reorganization of transport mechanisms leads to an increase angular velocity transfer efficiency and consequently enhanced drag.

In addition to bulk flow transition, particle settling also strongly affects near-wall dynamics. Light particles reduce viscous dissipation near the inner cylinder wall and slightly increase the near-wall velocity, consistent with drag reduction. In contrast, heavy particles significantly enhance near-wall dissipation and sharply decrease the azimuthal velocity near the wall with strong effective gravity. These effects result in a much thinner velocity boundary layer. These findings highlight the critical role of particle-induced changes to near-wall shear and turbulence in shaping the overall momentum transport.

These results offer new insights into particle-induced flow structure transitions in multiphase wall-bounded turbulence and offer strategies for flow control via tailored particle properties. Future studies may extend this framework to higher Reynolds numbers, or explore the effects of particle shape, and non-dilute conditions, thereby advancing the design of particle-laden flows for energy-efficient applications.

\section*{Acknowledgements}
This work is supported by NSFC Excellence Research Group Program for ‘‘Multiscale Problems in Nonlinear Mechanics’’ (No. 12588201), it was also supported by the Natural Science Foundation of China under grant nos. 12372219, 52450304, 12432011, 12422208, 12421002, and 12372220.

\section*{Declaration of interests}
The authors report no conflict of interest.

% \appendix
% \section{\label{Appendix}Derivation of the radial particle concentration profile} 

\bibliographystyle{jfm}
% % Note the spaces between the initials
\bibliography{literatur}

\begin{thebibliography}{47}
\expandafter\ifx\csname natexlab\endcsname\relax\def\natexlab#1{#1}\fi
\def\au#1{#1} \def\ed#1{#1} \def\yr#1{#1}\def\at#1{#1}\def\jt#1{\textit{#1}} \def\bt#1{#1}\def\bvol#1{\textbf{#1}} \def\vol#1{#1} \def\pg#1{#1} \def\publ#1{#1}\def\arxiv#1{#1}\def\org#1{#1}\def\st#1{\textit{#1}}

\bibitem[Assen {\em et~al.\/}(2022)Assen, Ng, Will, Stevens, Lohse \& Verzicco]{assen2022strong}
{\sc \au{Assen, M.~P.}, \au{Ng, C.~S.}, \au{Will, J.~B.}, \au{Stevens, R.~J.}, \au{Lohse, D.} \& \au{Verzicco, R.}} \yr{2022}  \at{Strong alignment of prolate ellipsoids in {T}aylor--{C}ouette flow}.  \jt{J.\ Fluid Mech.}  \bvol{935},  \pg{A7}.

\bibitem[Bakhuis {\em et~al.\/}(2018)Bakhuis, Verschoof, Mathai, Huisman, Lohse \& Sun]{bakhuis2018finite}
{\sc \au{Bakhuis, D.}, \au{Verschoof, R.~A.}, \au{Mathai, V.}, \au{Huisman, S.~G.}, \au{Lohse, D.} \& \au{Sun, C.}} \yr{2018}  \at{Finite-sized rigid spheres in turbulent {Taylor-Couette} flow: effect on the overall drag}.  \jt{J.\ Fluid Mech.}  \bvol{850},  \pg{246--261}.

\bibitem[Balachandar \& Eaton(2010)]{balachandar2010turbulent}
{\sc \au{Balachandar, S.} \& \au{Eaton, J.~K.}} \yr{2010}  \at{Turbulent dispersed multiphase flow}.  \jt{Annu.\ Rev.\ Fluid Mech.}  \bvol{42},  \pg{111--133}.

\bibitem[Baroudi {\em et~al.\/}(2020)Baroudi, Majji \& Morris]{baroudi2020effect}
{\sc \au{Baroudi, L.}, \au{Majji, M.~V.} \& \au{Morris, J.~F.}} \yr{2020}  \at{Effect of inertial migration of particles on flow transitions of a suspension {Taylor-Couette} flow}.  \jt{Phys.\ Rev.\ Fluids}  \bvol{5}~(11),  \pg{114303}.

\bibitem[Baroudi {\em et~al.\/}(2023)Baroudi, Majji, Peluso \& Morris]{baroudi2023taylor}
{\sc \au{Baroudi, L.}, \au{Majji, M.~V.}, \au{Peluso, S.} \& \au{Morris, J.~F.}} \yr{2023}  \at{{T}aylor--{C}ouette flow of hard-sphere suspensions: overview of current understanding}.  \jt{Philosophical Transactions of the Royal Society A}  \bvol{381}~(2243),  \pg{20220125}.

\bibitem[Bragg {\em et~al.\/}(2021)Bragg, Richter \& Wang]{bragg2021settling}
{\sc \au{Bragg, A.}, \au{Richter, D.} \& \au{Wang, G.}} \yr{2021}  \at{Settling strongly modifies particle concentrations in wall-bounded turbulent flows even when the settling parameter is asymptotically small}.  \jt{Phys.\ Rev.\ Fluids}  \bvol{6}~(12),  \pg{124301}.

\bibitem[Brandt \& Coletti(2022)]{brandt2022particle}
{\sc \au{Brandt, L.} \& \au{Coletti, F.}} \yr{2022}  \at{Particle-laden turbulence: progress and perspectives}.  \jt{Annu.\ Rev.\ Fluid Mech.}  \bvol{54}~(1),  \pg{159--189}.

\bibitem[Brauckmann \& Eckhardt(2013)]{brauckmann2013direct}
{\sc \au{Brauckmann, H.~J.} \& \au{Eckhardt, B.}} \yr{2013}  \at{Direct numerical simulations of local and global torque in {T}aylor--{C}ouette flow up to ${Re}= 30 000$}.  \jt{J.\ Fluid Mech.}  \bvol{718},  \pg{398--427}.

\bibitem[Carlotti \& Maggi(2021)]{carlotti2021experimental}
{\sc \au{Carlotti, S.} \& \au{Maggi, F.}} \yr{2021}  \at{Experimental techniques for characterization of particles in plumes of sub-scale solid rocket motors}.  \jt{Acta Astronautica}  \bvol{186},  \pg{496--507}.

\bibitem[Chouippe {\em et~al.\/}(2014)Chouippe, Climent, Legendre \& Gabillet]{chouippe2014numerical}
{\sc \au{Chouippe, A.}, \au{Climent, {\'E}.}, \au{Legendre, D.} \& \au{Gabillet, C.}} \yr{2014}  \at{Numerical simulation of bubble dispersion in turbulent {T}aylor-{C}ouette flow}.  \jt{Phys.\ Fluids}  \bvol{26}~(4).

\bibitem[Clarke \& Davoodi(2025)]{clarke2025movement}
{\sc \au{Clarke, A.} \& \au{Davoodi, M.}} \yr{2025}  \at{The movement of particles in {T}aylor--{C}ouette flow of complex fluids}.  \jt{J. Non-newton. Fluid.}  \bvol{335},  \pg{105354}.

\bibitem[Costa {\em et~al.\/}(2021)Costa, Brandt \& Picano]{costa2021near}
{\sc \au{Costa, P.}, \au{Brandt, L.} \& \au{Picano, F.}} \yr{2021}  \at{Near-wall turbulence modulation by small inertial particles}.  \jt{J.\ Fluid Mech.}  \bvol{922},  \pg{A9}.

\bibitem[Dash {\em et~al.\/}(2020)Dash, Anantharaman \& Poelma]{dash2020particle}
{\sc \au{Dash, A.}, \au{Anantharaman, A.} \& \au{Poelma, C.}} \yr{2020}  \at{Particle-laden {Taylor}--{Couette} flows: higher-order transitions and evidence for azimuthally localized wavy vortices}.  \jt{J.\ Fluid Mech.}  \bvol{903},  \pg{A20}.

\bibitem[Dong(2007)]{DONG2007Direct}
{\sc \au{Dong, S.}} \yr{2007}  \at{Direct numerical simulation of turbulent {T}aylor–{C}ouette flow}.  \jt{J.\ Fluid Mech.}  \bvol{587},  \pg{373--393}.

\bibitem[Eckhardt {\em et~al.\/}(2007)Eckhardt, Grossmann \& Lohse]{eckhardt2007torque}
{\sc \au{Eckhardt, B.}, \au{Grossmann, S.} \& \au{Lohse, D.}} \yr{2007}  \at{Torque scaling in turbulent {T}aylor--{C}ouette flow between independently rotating cylinders}.  \jt{J.\ Fluid Mech.}  \bvol{581},  \pg{221--250}.

\bibitem[Gao {\em et~al.\/}(2024)Gao, Wang \& Parsani]{gao2024drag}
{\sc \au{Gao, W.}, \au{Wang, M.} \& \au{Parsani, M.}} \yr{2024}  \at{Drag modulation by inertial particles in a drag-reduced turbulent channel flow with spanwise wall oscillation}.  \jt{Phys.\ Fluids}  \bvol{36}~(10).

\bibitem[Gatignol(1983)]{Gatignol_jmta_1983}
{\sc \au{Gatignol, R.}} \yr{1983}  \at{The {F}axén formulae for a rigid particle in an unsteady non-uniform {Stokes} flow}.  \jt{Journal de Mecanique Theorique et Appliquee}  \bvol{9}~(2),  \pg{143--160}.

\bibitem[Grossmann {\em et~al.\/}(2016)Grossmann, Lohse \& Sun]{grossmann2016high}
{\sc \au{Grossmann, S.}, \au{Lohse, D.} \& \au{Sun, C.}} \yr{2016}  \at{High-{Reynolds} number {Taylor-Couette} turbulence}.  \jt{Annu.\ Rev.\ Fluid Mech.}  \bvol{48},  \pg{53--80}.

\bibitem[Huisman {\em et~al.\/}(2013)Huisman, Scharnowski, Cierpka, K{\"a}hler, Lohse \& Sun]{huisman2013logarithmic}
{\sc \au{Huisman, S.~G.}, \au{Scharnowski, S.}, \au{Cierpka, C.}, \au{K{\"a}hler, C.~J.}, \au{Lohse, D.} \& \au{Sun, C.}} \yr{2013}  \at{Logarithmic boundary layers in strong {Taylor-Couette} turbulence}.  \jt{Phys.\ Rev.\ Lett.}  \bvol{110}~(26),  \pg{264501}.

\bibitem[Jiang {\em et~al.\/}(2025)Jiang, Lu, Wang, Meng, Shen \& Chong]{Jiang2025spatial}
{\sc \au{Jiang, H.}, \au{Lu, Z.-M.}, \au{Wang, B.-F.}, \au{Meng, X.~H.}, \au{Shen, J.} \& \au{Chong, K.~L.}} \yr{2025}  \at{Spatial distribution of inertial particles in turbulent {T}aylor--{C}ouette flow}.  \jt{J.\ Fluid Mech.}  \bvol{1006},  \pg{A2}.

\bibitem[Kim {\em et~al.\/}(2021)Kim, Oshima, Park \& Murai]{kim2021direct}
{\sc \au{Kim, S.}, \au{Oshima, N.}, \au{Park, H.~J.} \& \au{Murai, Y.}} \yr{2021}  \at{Direct numerical simulation of frictional drag modulation in horizontal channel flow subjected to single large-sized bubble injection}.  \jt{Int.\ J.\ Multiphase\ Flow}  \bvol{145},  \pg{103838}.

\bibitem[Lu {\em et~al.\/}(2005)Lu, Fern{\'a}ndez \& Tryggvason]{lu2005effect}
{\sc \au{Lu, J.}, \au{Fern{\'a}ndez, A.} \& \au{Tryggvason, G.}} \yr{2005}  \at{The effect of bubbles on the wall drag in a turbulent channel flow}.  \jt{Phys.\ Fluids}  \bvol{17}~(9).

\bibitem[Majji \& Morris(2018)]{majji2018inertial}
{\sc \au{Majji, M.~V.} \& \au{Morris, J.~F.}} \yr{2018}  \at{Inertial migration of particles in {Taylor-Couette} flows}.  \jt{Phys.\ Fluids}  \bvol{30}~(3).

\bibitem[Mathai {\em et~al.\/}(2020)Mathai, Lohse \& Sun]{mathai2020bubbly}
{\sc \au{Mathai, V.}, \au{Lohse, D.} \& \au{Sun, C.}} \yr{2020}  \at{Bubbly and buoyant particle--laden turbulent flows}.  \jt{Annu. Rev. Condens. Matter Phys.}  \bvol{11}~(1),  \pg{529--559}.

\bibitem[Maxey(1987)]{maxey1987gravitational}
{\sc \au{Maxey, M.~R.}} \yr{1987}  \at{The gravitational settling of aerosol particles in homogeneous turbulence and random flow fields}.  \jt{J.\ Fluid Mech.}  \bvol{174},  \pg{441--465}.

\bibitem[Maxey \& Riley(1983)]{maxey1983equation}
{\sc \au{Maxey, M.~R.} \& \au{Riley, J.~J.}} \yr{1983}  \at{Equation of motion for a small rigid sphere in a nonuniform flow}.  \jt{Phys.\ Fluids}  \bvol{26}~(4),  \pg{883--889}.

\bibitem[Murai {\em et~al.\/}(2005)Murai, Oiwa \& Takeda]{murai2005bubble}
{\sc \au{Murai, Y.}, \au{Oiwa, H.} \& \au{Takeda, Y.}} \yr{2005} Bubble behavior in a vertical {T}aylor-{C}ouette flow.  \bt{In {\em Journal of Physics: conference series\/}}, ,  \vol{vol.~14},  \pg{p. 143}. IOP Publishing.

\bibitem[Naumann \& Schiller(1935)]{naumann1935drag}
{\sc \au{Naumann, Z.} \& \au{Schiller, L.}} \yr{1935}  \at{A drag coefficient correlation}.  \jt{Z. Ver. Deutsch. Ing}  \bvol{77}~(318),  \pg{e323}.

\bibitem[Ni(2024)]{ni2024deformation}
{\sc \au{Ni, R.}} \yr{2024}  \at{Deformation and breakup of bubbles and drops in turbulence}.  \jt{Annu.\ Rev.\ Fluid Mech.}  \bvol{56}~(1),  \pg{319--347}.

\bibitem[Ostilla {\em et~al.\/}(2013)Ostilla, Stevens, Grossmann, Verzicco \& Lohse]{ostilla2013optimal}
{\sc \au{Ostilla, R.}, \au{Stevens, R.~J.}, \au{Grossmann, S.}, \au{Verzicco, R.} \& \au{Lohse, D.}} \yr{2013}  \at{Optimal {T}aylor--{C}ouette flow: direct numerical simulations}.  \jt{J.\ Fluid Mech.}  \bvol{719},  \pg{14--46}.

\bibitem[Ostilla-M{\'o}nico {\em et~al.\/}(2014)Ostilla-M{\'o}nico, van~der Poel, Verzicco, Grossmann \& Lohse]{ostilla2014boundary}
{\sc \au{Ostilla-M{\'o}nico, R.}, \au{van~der Poel, E.~P.}, \au{Verzicco, R.}, \au{Grossmann, S.} \& \au{Lohse, D.}} \yr{2014}  \at{Boundary layer dynamics at the transition between the classical and the ultimate regime of {Taylor-Couette} flow}.  \jt{Phys.\ Fluids}  \bvol{26}~(1).

\bibitem[Russo {\em et~al.\/}(2014)Russo, Kuerten, van~der Geld \& Geurts]{russo2014water}
{\sc \au{Russo, E.}, \au{Kuerten, J.~G.}, \au{van~der Geld, C.~W.} \& \au{Geurts, B.~J.}} \yr{2014}  \at{Water droplet condensation and evaporation in turbulent channel flow}.  \jt{J.\ Fluid Mech.}  \bvol{749},  \pg{666--700}.

\bibitem[Sippola {\em et~al.\/}(2018)Sippola, Kolehmainen, Ozel, Liu, Saarenrinne \& Sundaresan]{sippola2018experimental}
{\sc \au{Sippola, P.}, \au{Kolehmainen, J.}, \au{Ozel, A.}, \au{Liu, X.}, \au{Saarenrinne, P.} \& \au{Sundaresan, S.}} \yr{2018}  \at{Experimental and numerical study of wall layer development in a tribocharged fluidized bed}.  \jt{J.\ Fluid Mech.}  \bvol{849},  \pg{860--884}.

\bibitem[Spandan {\em et~al.\/}(2016{\natexlab{{\em a\/}}})Spandan, Lohse \& Verzicco]{spandan2016deformation}
{\sc \au{Spandan, V.}, \au{Lohse, D.} \& \au{Verzicco, R.}} \yr{2016{\natexlab{{\em a\/}}}}  \at{{Deformation and orientation statistics of neutrally buoyant sub-Kolmogorov ellipsoidal droplets in turbulent {Taylor-Couette} flow}}.  \jt{J.\ Fluid Mech.}  \bvol{809},  \pg{480--501}.

\bibitem[Spandan {\em et~al.\/}(2016{\natexlab{{\em b\/}}})Spandan, Ostilla-M{\'o}nico, Verzicco \& Lohse]{spandan2016drag}
{\sc \au{Spandan, V.}, \au{Ostilla-M{\'o}nico, R.}, \au{Verzicco, R.} \& \au{Lohse, D.}} \yr{2016{\natexlab{{\em b\/}}}}  \at{{Drag reduction in numerical two-phase {Taylor-Couette} turbulence using an {Euler-Lagrange} approach}}.  \jt{J.\ Fluid Mech.}  \bvol{798},  \pg{411--435}.

\bibitem[Su {\em et~al.\/}(2024)Su, Yi, Zhao, Wang, Xu, Wang \& Sun]{su2024numerical}
{\sc \au{Su, J.}, \au{Yi, L.}, \au{Zhao, B.}, \au{Wang, C.}, \au{Xu, F.}, \au{Wang, J.} \& \au{Sun, C.}} \yr{2024}  \at{Numerical study on the mechanism of drag modulation by dispersed drops in two-phase {T}aylor--{C}ouette turbulence}.  \jt{J.\ Fluid Mech.}  \bvol{984},  \pg{R3}.

\bibitem[Sugiyama {\em et~al.\/}(2008)Sugiyama, Calzavarini \& Lohse]{sugiyama2008microbubbly}
{\sc \au{Sugiyama, K.}, \au{Calzavarini, E.} \& \au{Lohse, D.}} \yr{2008}  \at{Microbubbly drag reduction in {T}aylor--{C}ouette flow in the wavy vortex regime}.  \jt{J.\ Fluid Mech.}  \bvol{608},  \pg{21--41}.

\bibitem[Taylor(1923)]{taylor1923viii}
{\sc \au{Taylor, G.~I.}} \yr{1923}  \at{Viii. {Stability} of a viscous liquid contained between two rotating cylinders}.  \jt{Phil. Trans. R. Soc. Lond. Series A}  \bvol{223}~(605-615),  \pg{289--343}.

\bibitem[Tsai(2022)]{Tsai_taml_2022}
{\sc \au{Tsai, S.-T.}} \yr{2022}  \at{Sedimentation motion of sand particles in moving water {(I)}: The resistance on a small sphere moving in non-uniform flow}.  \jt{Theor. Appl. Mech. Lett.}  \bvol{12}~(6),  \pg{100392}.

\bibitem[Verzicco \& Orlandi(1996)]{verzicco1996finite}
{\sc \au{Verzicco, R.} \& \au{Orlandi, P.}} \yr{1996}  \at{A finite-difference scheme for three-dimensional incompressible flows in cylindrical coordinates}.  \jt{J. Comput. Phys.}  \bvol{123}~(2),  \pg{402--414}.

\bibitem[Wang {\em et~al.\/}(2022{\natexlab{{\em a\/}}})Wang, Yi, Jiang \& Sun]{wang2022finite}
{\sc \au{Wang, C.}, \au{Yi, L.}, \au{Jiang, L.} \& \au{Sun, C.}} \yr{2022{\natexlab{{\em a\/}}}}  \at{How do the finite-size particles modify the drag in {T}aylor--{C}ouette turbulent flow}.  \jt{J.\ Fluid Mech.}  \bvol{937},  \pg{A15}.

\bibitem[Wang {\em et~al.\/}(2022{\natexlab{{\em b\/}}})Wang, Yi, Jiang \& Sun]{wang_yi_jiang_sun_2022}
{\sc \au{Wang, C.}, \au{Yi, L.}, \au{Jiang, L.-F.} \& \au{Sun, C.}} \yr{2022{\natexlab{{\em b\/}}}}  \at{How do the finite-size particles modify the drag in {Taylor–Couette} turbulent flow}.  \jt{J.\ Fluid Mech.}  \bvol{937},  \pg{A15}.

\bibitem[Wang {\em et~al.\/}(2021)Wang, Xu \& Zhao]{wang2021turbulence}
{\sc \au{Wang, Z.}, \au{Xu, C.-X.} \& \au{Zhao, L.}} \yr{2021}  \at{Turbulence modulations and drag reduction by inertialess spheroids in turbulent channel flow}.  \jt{Phys.\ Fluids}  \bvol{33}~(12).

\bibitem[Wereley \& Lueptow(1999)]{wereley1999inertial}
{\sc \au{Wereley, S.~T.} \& \au{Lueptow, R.~M.}} \yr{1999}  \at{Inertial particle motion in a {T}aylor {C}ouette rotating filter}.  \jt{Phys.\ Fluids}  \bvol{11}~(2),  \pg{325--333}.

\bibitem[Zhang \& Zhou(2023)]{zhang2023unveiling}
{\sc \au{Zhang, H.} \& \au{Zhou, Y.-H.}} \yr{2023}  \at{Unveiling the spectrum of electrohydrodynamic turbulence in dust storms}.  \jt{Nat.\ Commun.}  \bvol{14}~(1),  \pg{408}.

\bibitem[Zhang {\em et~al.\/}(2020)Zhang, Wang \& Wan]{zhang2020euler}
{\sc \au{Zhang, X.}, \au{Wang, J.} \& \au{Wan, D.}} \yr{2020}  \at{Euler--{L}agrange study of bubble drag reduction in turbulent channel flow and boundary layer flow}.  \jt{Phys.\ Fluids}  \bvol{32}~(2).

\bibitem[Zhang {\em et~al.\/}(2025)Zhang, Fan, Su, Xi \& Sun]{zhang2025global}
{\sc \au{Zhang, Y.-B.}, \au{Fan, Y.}, \au{Su, J.}, \au{Xi, H.-D.} \& \au{Sun, C.}} \yr{2025}  \at{Global drag reduction and local flow statistics in {T}aylor--{C}ouette turbulence with dilute polymer additives}.  \jt{J.\ Fluid Mech.}  \bvol{1002},  \pg{A33}.

\end{thebibliography}

\end{document}